\documentclass[journal]{IEEEtran}
\ifCLASSINFOpdf
\else
\fi
\usepackage{amsmath,amssymb,graphicx}
\usepackage{xcolor}
\usepackage{url}

\begin{document}

\title{Ultra-narrow spectral response of a \\ hybrid plasmonic-grating sensor}

\author{Mahmoud H. Elshorbagy, Alexander Cuadrado, Braulio Garc´\'{\i}a-C\'{a}mara, Ricardo Vergaz, \\
Jos\'{e} Antonio G\'{o}mez-Pedrero, and Javier Alda 
\thanks{M. Elshorbagy, A. Cuadrado, J. A. G\'{o}mez-Pedrero and J. Alda are with the Applied Optics Complutense Group. University Complutense of Madrid. 
Faculty of Optics and Optometry. Av. Arcos de Jal\'{o}n, 118. 28037 Madrid, Spain.}
\thanks{M. Elshorbagy, B. Garc\'{\i}a C\'{a}mara, and R. Vergaz are with the Group of Displays and Photonic Applications (GDAF), Carlos III University  of Madrid. Department of Electronic Technology. Av. de la Universidad, 30. 
28911 Legan\'{e}s. Spain.}
\thanks{M. Elsorhbagy is also with the Physics Department, Faculty of Science, Minia University. University campus, 61519, El-Minya, Egypt.}
\thanks{A. Cuadrado is also with 
Escuela Superior de Ciencias Experimentales y Tecnolog\'{\i}a. 
  Universidad Rey Juan Carlos, 28933 Madrid, Spain.}
\thanks{J. Alda (javier.alda@ucm.es) is the corresponding author}
}


\maketitle


\begin{abstract}

We configure and analyze   a nanostructured device that hybridizes  grating modes and surface plasmon resonances. 
The model uses an effective index of refraction that considers the volume fraction of the involved materials, and the propagation depth of the plasmon through the structure.
Our geometry  is an  extruded low-order diffraction grating made of dielectric nano-triangles.
Surface plasmon resonances are excited at a metal/dielectric interface, which is separated from the analyte by a high-index dielectric layer.
The optical performance of the refractometric sensor is highly competitive in sensitivity and figure of merit (FOM) because of the the  ultra-narrow spectral response (below 0.1 nm). Moreover, it is operative within a wide range of the index of refraction (from 1.3 till 1.56), and also works under normal incidence conditions.

\end{abstract}

\begin{IEEEkeywords}
nanophotonics, plasmonics, optical sensors.
\end{IEEEkeywords}

\IEEEpeerreviewmaketitle

\textcopyright 2019 IEEE. Personal use is permitted, but republication/redistribution requires IEEE permission. Digital Object Identifier DOI: 10.1109/JSEN.2019.2960556
\url{https://ieeexplore.ieee.org/abstract/document/8936420}

\section{Introduction}

The generation of surface plasmon resonances using waveguides and diffraction gratings has been extensively studied, and successfully applied to advanced photonic devices in the last decade
\cite{Yang2018Characteristics,Hayashi_2015}. 
These systems are fast, efficient, and portable; and their optical response is customized and adapted to  energy harvesting \cite{BRADY2019197,REZAEI2019325}, material characterization and quality testing \cite{LEIGH2019885,Alahnomi2019determaination,SHAHVAR2019578}, environmental and chemicals sensing \cite{ABRAHAM2019125,OZDEMIR201954}, and  biomedical applications \cite{MANVI201957,Barbhaiya2019}.

Currently, device's improvement  relies on the extension of its response to a wider dynamic range for multi-sensing applications, and on a better performance for dedicated sensors in specific applications \cite{AHMAD2012382,Elshorbagy2017,Elshorbagy2017prism,Elshorbagy2019}.
Optical sensors outperform previous techniques  through the exploitation of the characteristics of surface plasmon resonances (SPR). 
Certain nanophotonic structures with  ultra-narrow spectral response (in the nanometric range) allow for  high-resolution sensing and imaging devices  \cite{Liu2019,REN2019308,WU201925}. Other photonic structures are intended for wide spectral responses, as in energy harvesting devices, where they improve efficiency for low cost devices  \cite{Atwater2010,ELSHORBAGY2017130,LEE2009416}. 

A surface plasmon wave is excited at the interface between metal/dielectric media when the incoming light satisfies the wave matching condition at that interface \cite{Elshorbagy2017}. 
The Kretschmann and Otto configurations were proposed to fulfill the matching conditions that generate the SPR appropriate for sensing applications \cite{otto1967theory,Kretschmann1968}.  
When light is efficiently coupled to the modes supported by the geometric and material arrangements of the device, not only is the spectral response narrow, but also the field is spatially localized and enhanced. 
This is key for sensing applications because the interaction of light is strongly affected by the optical characteristics of the exposed media \cite{Elshorbagy2019}.
The narrow spectral response, due to SPR excitation, becomes even narrower if these resonances interact with 
guided or grating modes to generate hybrid resonances \cite{Lee2010,Elshorbagy2017,Elshorbagy2017prism,Elshorbagy2019}.

In optical sensing,  several parameters inform us  how to adapt each configuration for a given application. 
These parameters include: sensitivity, figure of merit (FOM), limit of detection, dynamic range, resolution, and linearity. 
Furthermore, the competitiveness of a sensor also depends on its fabrication feasibility, and the simplicity of operation \cite{Piliarik2009}. 
Sensitivity, FOM, detection limit, and resolution characterize how the sensor's response changes when the measured property varies (some of these parameters use the full-width-at-half-maximum, FWHM, of the spectral or angular response).
Recent advances in optical sensors have reported extremely good values of the mentioned parameters, with values of FWHM below 1 nm \cite{Wu2016,Feng:18, liu2018plasmonic},  and FOM up to 630 RIU$^{-1}$.

This contribution focus on the combination of a low-order diffraction grating and a metal/dielectric interface where the SPR is generated. 
Therefore, the period of the grating, $P$, is similar to the wavelength $\lambda$. 
In the proposed configuration, 
SPRs are excited in reflection at the interface between a thick metal layer and a dielectric buffer. 
On top of this dielectric layer, a periodic grating made of dielectric nano-triangles route the light towards the metal/dielectric interface.
This triangular relief also triggers funneling and guiding mechanisms that improve the performance of the devices \cite{Elshorbagy2017prism,elshorbagy_jfpenergy_2017}.
 Light impinges the structure from the analyte side that is located above the dielectric grating.
In this case, the excitation of a SPR  depends 
 on the index of refraction of the dielectric in contact with the metal, on the grating parameters, 
and on the effective index obtained by the combination of the dielectric slab, the material of the nano-triangles, and the index of refraction of the analyte.
This combination produces a multi-resonance behavior related with low-order diffraction orders. 
This hybridization has been already presented in recent contributions \cite{pi_ieeephotj206, li2019appphyslet}, but the ultra-narrow  hybridized resonances reported in this paper generate higher sensitivity and FOM, that expands over a wide dynamic range with reasonable performance.
The device operates under normal incidence conditions, easing its integration for both illumination and signal retrieval systems, and allowing to locate it at the tip of an optical fiber. 
A challenging application for the device presented here would be the measurement of refractive index change of the analyte when the volume of analyte is low (nano- or picoliters). This could be advantageously used in sensing changes of the tear film, with potential application in the diagnostics of dry-eye syndrome \cite{Khalil_tearsensor_2004,benito_tear_jcrs_11,kottaiyan_tear_ocusur_2012}.

This contribution is organized as follows. In section \ref{sec:model} we describe the physical mechanisms at play in this device, propose a feasible structure showing the reflectance dip of interest, and optimize it by selecting the geometric parameters, and the materials for improved performance. We analyze the device and characterize it as a refractometric sensor in section \ref{sec:deviceanalysis}. Finally, the main conclusions of the paper are presented in section \ref{sec:conclusions}.
 
\section{Model, design, and optimization}

\label{sec:model}

\begin{figure}[h!]
\centering
  \includegraphics[width=0.95\columnwidth]{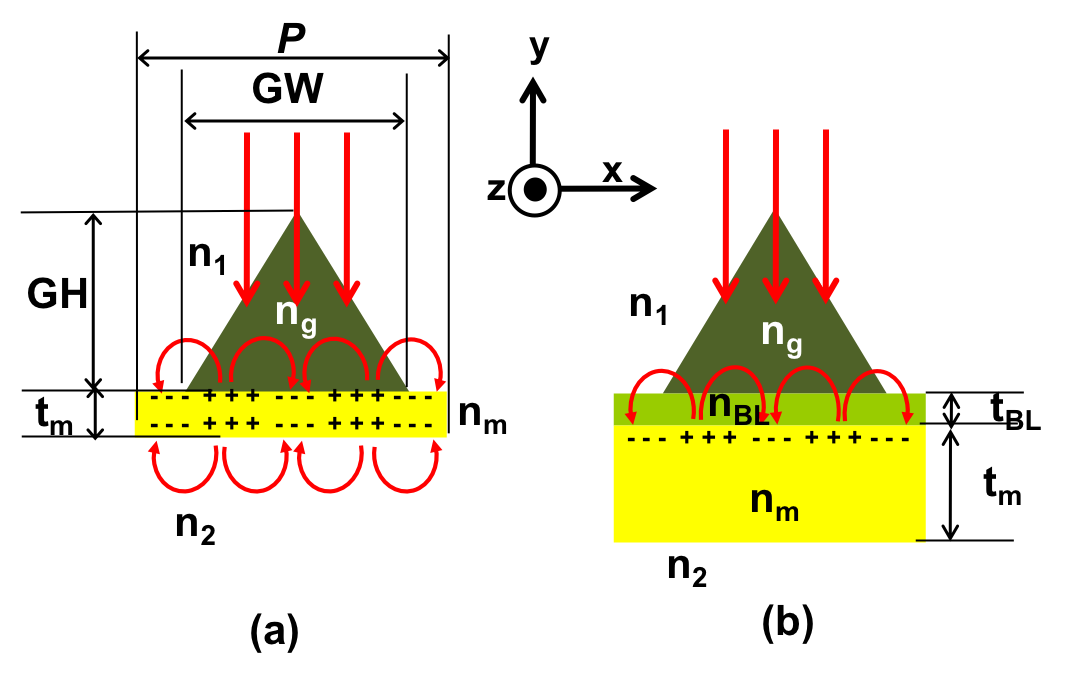}
  \caption{Geometrical and material cross sections of two configurations for the generation of SPRs combined with a low-order dielectric grating. 
 This grating has a period $P$, and is carried out with dielectric nanotriangles. Theses nanotriangles have a width GW, and a height GH, and they are made with a dielectric material having an index of refraction $n_g$. The metal layer, having an index $n_m$, has a thickness $t_m$.  The configuration in (b) has an additional buffer layer with an index of refraction $n_{\rm BL}$, and a thickness $t_{\rm BL}$. 
  Light is coming from top in both configurations, and the plasmonic resonance can be generated on both sides of the metal layer if this is thin enough (case (a)) or only at the top metal/dielectric interface (case (b)). The substrate supporting these structures, having an index of refraction $n_s$, is placed on top, $n_1=n_s$ (case (a)), or at the bottom, $n_2=n_s$ (case (b)). Meanwhile, the analyte, with an index of refraction $n_a$, is located at the bottom, $n_2=n_a$ (case(a)), or on top, $n_1=n_a$ (case (b)). The geometrical parameters of the structure are depicted in the $XY$ plane, being $Z-$axis the extrusion axis of the 3D nanostructure. }
  \label{fig:gratings}
\end{figure}

In this paper, we analyze the performance of nanophotonic structures that combine low-order diffraction grating modes and SPRs. Figure \ref{fig:gratings} shows two cases with hybrid behavior. 
A diffraction grating of period $P$ along $X-$direction, and made of dielectric nano-triangles with an index of refraction $n_g$, is placed on top of a material structure that may take several configurations described as a cross section on the $XY$ plane.
We will extrude this geometry along the $Z-$axis.  
Surface plasmons are generated at a metal/dielectric interface where the real part of the dielectric permittivity of the metal ($\varepsilon_m$) is negative, meanwhile the index of refraction of the dielectric remains real and positive. 
Our device is illuminated from top through a medium of index of refraction $n_1$. 
Depending on the configuration, the analyte (having an index of refraction $n_a$) can be placed on top of the nano-triangles ($n_1=n_a$), or at the bottom of the structure, interfacing with an index of refraction $n_2=n_a$ (see Fig \ref{fig:gratings}.a). 
If the metal is thick enough (greater than some few penetration depths within the metal), the plasmon resonance is suppressed at the bottom of the structure, and a dielectric buffer layer (having a high index of the refraction, $n_{\rm BL}$) between the metal and the nano-triangles is used to promote and enhance resonances on top of the metal layer 
(see Fig. \ref{fig:gratings}.b). 
The substrate is placed on top for design in Fig. 
\ref{fig:gratings}.a, and at the bottom in Fig. \ref{fig:gratings}.b.
In the second case, it is the effective medium that combines the grating material, the analyte and the buffer layer what effectively controls the SPR generation. That provides the designer with a deeper control of the desired effect, due to the higher contrast of the refractive indices in this configuration.
In any case, the excitation of a SPR at a metal/dielectric interface coupled to a diffraction grating is driven 
by the wavevector matching condition \cite{Abutoama2015},
 \begin{equation}
\vec{k}_{0x} + \vec{k}_{{\rm grating}}= \vec{k}_{{\rm spp}}                
,\label{eq:wavevector}
\end{equation}
where $\vec{k}_{0x}$ is the $x$-component of the wavevector of the incoming wavefront; $\vec{k}_{{\rm grating}}=2\pi M/ P$ is the grating wavevector, being $P$ the grating period, and $M$ the diffraction order; and $\vec{k}_{{\rm spp}}$ is the surface plasmon wavevector generated at the interface. If we include the angle of incidence in the wavevector matching conditions, it becomes \cite{Abutoama2015}:
\begin{equation}
n_{1} k_{0} \sin\theta \pm \frac{2\pi M}{P} = k_{0} \sqrt{\frac{\varepsilon_{m} \varepsilon_{{\rm eff}}}{\varepsilon_{m} +\varepsilon_{{\rm eff}}}}                             
,\label{eq:SPR}
\end{equation}
where $k_{0}$ is the wavevector in vacuum, $\theta$ is the angle of incidence of the incoming wave (in our case $\theta=0^\circ$ and it is not depicted in Fig. \ref{fig:gratings}), 
$\varepsilon_{m}$ is the real part of the complex electric permittivity of the metal, and 
$\varepsilon_{\rm eff}=n_{\rm eff}^2$ is the dielectric permittivity of an effective medium in contact with the metal. 
The use of an effective medium is needed because of the presence of
several materials in the surroundings of the metal interface: the dielectric thin layer, the dielectric triangular grating, and the analyte. 
Usually,  the effective index of refraction is 
calculated with 
the so-called homogenization method \cite{RYTOV1956}. In this method, each material contributes to the effective refractive index proportionally  to its volume fraction within the interaction volume of the plasmon (i.e., close to the metal/dielectric surface). 
We parameterize the reach of the surface plasmon with its decay length.
After applying Eq. (\ref{eq:SPR}) to the case of normal incidence conditions, $\theta=0^\circ$, and knowing that $k_0=2 \pi/\lambda$, 
the wavelength of the resonance is given as:
\begin{equation}
\lambda_{{\rm res}} = \frac{P}{M} \sqrt{\frac{\varepsilon_{m} \varepsilon_{{\rm eff}}}{\varepsilon_{m} +\varepsilon_{{\rm eff}}}}                             
.\label{eq:lambres}
\end{equation}

The structure  depicted in Fig. \ref{fig:gratings}.a 
 couples radiation towards a metallic thin layer. 
 The thickness of this layer, $t_m$, is thin enough ($t_m \simeq 45$ nm) to generate SPR in both sides of the metallic slab.
Therefore, the analyte can be placed on top and/or at the bottom of the structure.

When the metal layer is thick enough to prevent the excitation of the SPR at the bottom interface, we find the configuration shown in Fig. \ref{fig:gratings}.b. 
This configuration has an additional thin dielectric buffer layer  (being $n_{\rm BL}$ its index of refraction), with a thickness $t_{{\rm BL}}$ small compared to the decay length of the surface plasmon. This material appears at the wavevector matching condition for those SPR at the top of the metal layer included in the calculation of the effective index of refraction as a combination of the indices of refraction of three materials: buffer layer ($n_{\rm BL}$), nano-triangles ($n_g$), and analyte ($n_a$). 
This dielectric buffer layer also controls the coupling between the diffracted wave and the SPR mode, and improves the final response of the device \cite{Elshorbagy2017prism}.
In this case, SPRs are only excited on top of the metal layer.
The calculation of the effective index of refraction requires the estimation of the spatial extension of the plasmon measured from the top surface of the metal. 
This length is wavelength dependent and will be extracted from the simulations. 
Moreover, the presence of the diffraction grating generates low-order diffraction modes that interact with the structures. The model described in Eq. (\ref{eq:lambres}) explains the observed behavior. 
Furthermore, the combination of the grating lobes and the SPR leads, after optimization, to an enhancement in the device's sensitivity, a narrower spectral response, and therefore a larger FOM.

The electromagnetic behavior of the device is numerically analyzed and tested using COMSOL Multiphysics.

\subsection{Proposed structure}
\label{sec:proposedstructure}

In this section we focus on the design shown in Fig. \ref{fig:gratings}.b where the metal layer thickness prevents the excitation of SPR at the bottom interface of the metal layer. The analyte medium is on top of the grating ($n_1=n_a$), and plasmons are generated at the upper metal/dielectric interface. The combination of the high-index buffer layer and the presence of the dielectric grating produces a hybrid resonance with ultra-narrow spectral response.
 The geometrical and material arrangement is described in table \ref{tab:parameters}, organized from bottom to top, having SiO$_2$ as substrate and water as superstrate.  
The values of the index of refraction for the given materials are extracted from references \cite{JohnsonChristy1972}, \cite{AdachiSadao1989}, and \cite{Luke2015} for Ag, GaP, and Si$_3$N$_4$, respectively. The index of refraction of dielectrics is purely real, and their imaginary parts are neglected within the studied spectral range.

\begin{table}[h!]
\begin{center}
\caption{Geometrical parameters and materials for the proposed device. 
  \label{tab:parameters}}
\begin{tabular}{ccc}
\hline  
Material &  Dimension and shape &  $n(\lambda=1.4\mu$m) 
\\  \hline 
SiO$_2$  & $\infty$ (substrate) & $n_s=1.45$
\\ \hline 
Ag  & $t_m=200$ nm    (thin film) & $n_m=0.13065 - i 10.156$ 
\\ \hline  
GaP  & $t_{\rm BL}= 100$ nm   (thin film)  & $n_{\rm BL}=3.1369$ 
\\ \hline 
Si$_3$N$_4$  & Width GW=610 nm  & $n_g=2.0005$ \\ 
			  & Height GH=1070 nm  &  \\
 			  & Period, $P$= 1000 nm  &  
\\ \hline 
H$_2$O  & $\infty$  superstrate (analyte)  & $n_a=1.33$
\\ \hline 
\end{tabular}
\end{center}
\end{table}

The  SiO$_2$  substrate could be replaced by other materials, even plastic, because it 
does not influence the optical behavior.
For the grating and buffer layer, the selection of the dielectric materials is
key to obtain a high-contrast of the index of refraction. 
Our grating is made of Si$_3$N$_4$ with an index of refraction $n_g \simeq 2$ at the operating wavelength ($\lambda \in [1.35 - 1.55] \mu$m). 
This value is large compared with the index of refraction of the analyte. In this design, we take water as the analyte. We vary the index of refraction  between $1.30$ and $1.56$ to consider for any additions, variations, or even if water is replaced by other  bio-chemical media  \cite{Jin_2006,Mulloni2000}.
As we will show, both the triangular shape of the grating relief, and
the contrast in the index of refraction between the analyte and the grating enhance the funneling of light towards the dielectric/metal interface where the SPR is generated. 
This effect is also favored by the high-aspect ratio of the grating relief,  and also narrows the spectral response of the structure.
The buffer  must be a transparent and high-index material: we use GaP as starting point.  
Later on this  paper, we will analyze the importance of 
the thickness and material selection of
this buffer layer to  optimize the performance of the system.

\begin{figure}[h!]
\centering
  \includegraphics[width=0.80\columnwidth]{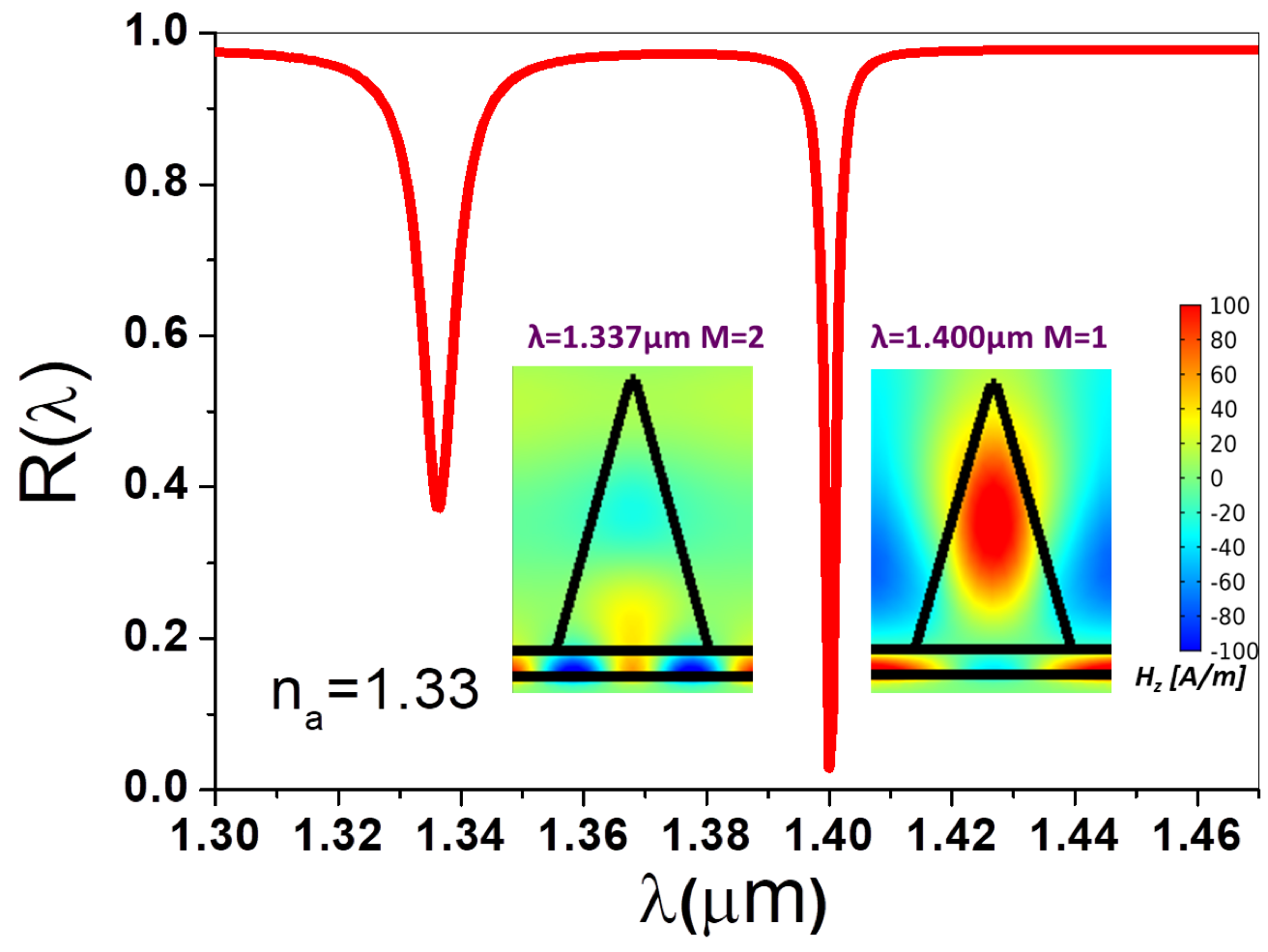}
  \caption{Spectral response  of the proposed structure. The insets represent the fields for $M=1$, and $M=2$ (the colorbar is the same for both insets).}
  \label{fig:response}
\end{figure}

This sensor can be 
easily implemented in an optical setup. 
We start with collimated light exiting from an optical fiber  and directed towards a measurement cell where the analyte fills the volume between the collimation optics and the sensing surface. 
The spectrum of the reflected light carries information about the analyte and  is collected by the same optical system towards detection.
We consider that the device is illuminated from the top with a TM polarized plane wave, under normal incidence conditions. 
In our simulation we have set an amplitude of 1 A/m for the magnitude of the magnetic field of the incoming wave. Then, the magnetic field map can be interpreted as a field enhancement map.  Under these conditions, we have calculated the spectral reflectance, $R(\lambda)$, that is presented in Fig. \ref{fig:response}. $R(\lambda)$ contains the mirror reflectance (0$^{\rm th}$ order) plus the low-order diffraction lobes ($\pm 1 ^{\rm st}$ orders).
The lower it is, the higher the power of the generated SPR, and thus the higher the sensitivity.
In this figure, when $n_a=1.33$, we see two minima located at $\lambda$=1.337 $\mu$m and 1.4 $\mu$m.
They result from the coupling of diffractive orders to the SPR. 
The first one ($\lambda=1.337 \mu$m) corresponds with a mode that is mostly confined in the buffer layer and has very little exposure to the analyte (see inset in Fig. \ref{fig:response}). 
However, the second minimum ($\lambda=1.4 \mu$m) shows a large amplitude both within the grating material and the analyte interspace, also beyond the triangular tips.

At this point in the simulation, we use  Eq. (\ref{eq:lambres})  to find the effective index of refraction (or effective permittivity). 
First, we  determine the diffraction order, $M$, of the resonance. This order can be obtained after analyzing the field distribution within a period of the grating. The insets in Fig. \ref{fig:response} show  periodic field amplitude distributions that repeat once within the period for $\lambda=1.400 \mu$m, meaning $M=1$, and twice for $\lambda=1.337 \mu$m, meaning $M=2$. 
After substituting in Eq. (\ref{eq:lambres}) the optical constant of the metal, Ag, and the value of the period, $P=1\mu$m,  we find an
effective index of refraction $n_{\rm eff}(\lambda=1.4 \mu{\rm m} )= 1.3869$,
and $n_{\rm eff}(\lambda=1.337 \mu{\rm m} )= 2.5778$ (where $n_{\rm eff}^2= \epsilon_{\rm eff}$).
These values of the effective index are given by a mixing model that includes the volume fractions of the materials of the arrangement. 
This volume fraction is obtained by considering
the amount of the volume of each material present in the unit cell (with period $P$) up to an arbitrary height referenced to the metal/dielectric interface, $h_{\rm eff}$.
When doing this calculation we find 
the height that fits the effective index values at each wavelength previously evaluated from Eq. (\ref{eq:lambres}):
 $h_{\rm eff}(\lambda=1.400 \mu{\rm m})=7.039 \mu$m, 
 and $h_{\rm eff}(\lambda=1.337 \mu{\rm m})=0.164 \mu$m. 
 These height values can be related to the extent of the region where the field interacts, as we will see when evaluating the field enhancement maps in the next subsection \ref{sec:optimization}.
 The feasibility of these values supports the previously presented mode of the hybridization mechanism between SPR and low-order diffraction orders.

\subsection{Optimization}
\label{sec:optimization}

As customary, we compare the performance of optical sensors by the sensitivity and the figure of merit 
\cite{HOMOLA199916,Cennamo2013,Spackova2016}.
Usually, when light efficiently couples to SPR, the field enhancement is larger and the reflectance reaches a minimum value. This is why we will pay attention to this enhancement, quantitatively expressed as FE$(x,y,z)=H(x,y,z)/H_{\rm in}$ (where $H(x,y,z)$ is the magnetic field at an arbitrary  location within the volume, and $H_{\rm in}$ is the amplitude of the magnetic field of the input wavefront), and also to the spectral reflectivity $R(\lambda)$. These two parameters can be combined
in a merit function, ${\rm MF}$, that is defined as:
\begin{equation}
{\rm MF}={\rm FE} (1-R)
.\label{eq:merit}
\end{equation}
The field enhancement parameter, FE, is defined here as the maximum value of FE$(x,y,z)$ within the analyte volume.
Reflectivity characterizes  the behavior of the structure globally at the wavelength of resonance, $R=R(\lambda_{\rm res})$. This merit function highlights hot spots in the field map, and distinguishes the optimum coupling of the incoming radiation to the SPR (in this case, as far as transmittance is negligible, a minimum reflectance means a maximum absorption).

\begin{figure}[h!]
\centering
  \includegraphics[width=0.90\columnwidth]{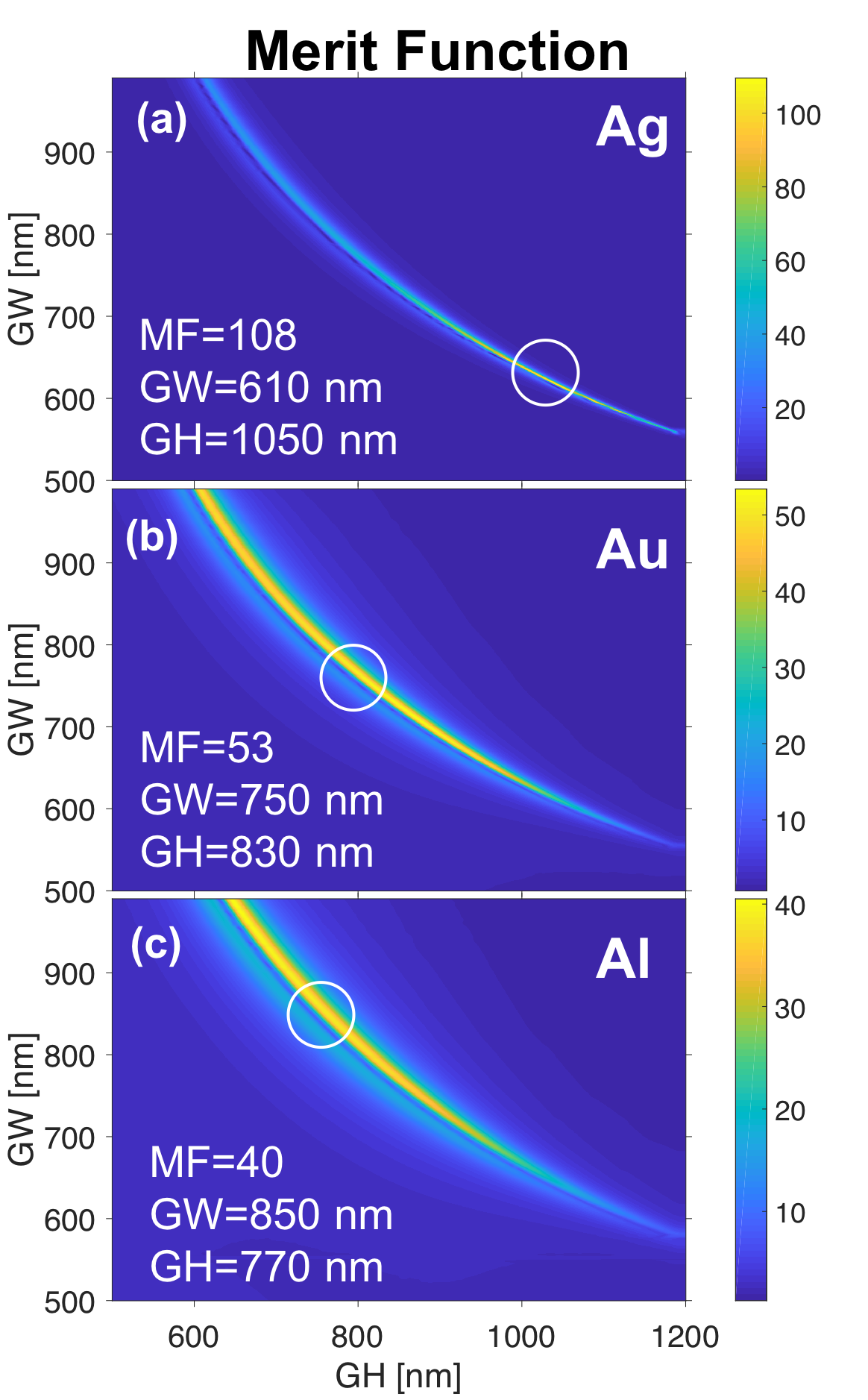}
  \caption{Maps of the merit function (Eq. (\ref{eq:merit})) for three metals as a function of the geometrical parameters of the grating, GW and GH. (a) is for silver, (b) is for gold, and (c) is for aluminum. The grating period is $P=1\mu$m, $\lambda=1.4\mu$m and $n_a=1.33$. The white circles are centered around the maximum value of the MF. The colormap expands along different ranges for each metal.}
  \label{fig:optimzation}
\end{figure}

The goal of the optimization is to set the geometrical and material parameters that maximize this merit function (Eq. \ref{eq:merit}).
We first analyze the effect of the geometry of the dielectric grating and the metal selection, and then we will move to optimize the thickness and material choice of the buffer layer.
Figure \ref{fig:optimzation} shows how the grating geometry  (grating width, GW, and grating height, GH) affects MF for fixed values of the wavelength ($\lambda=1.4 \mu$m) and the period of the grating  ($P=1 \mu$m).
Actually, changing $P$ moves the  wavelength of resonance linearly (as larger $P$, larger is $\lambda_{\rm res}$), and allows 
to tune the design for a given wavelength of interest. 
The three optimization maps  consider three possible choices for the metal: silver, gold, and aluminum. 
We select these materials because they are widely used in plasmonic sensors  \cite{MITSUSHIO2006296}. 
The dependence of MF with respect to GW and GH for the three metals is similar.
Ag gives the largest value of MF, and Al the smallest (see the different range in the color bar on the right of the maps in Fig. \ref{fig:optimzation} for each metal).

\begin{figure}[h!]
\centering
  \includegraphics[width=0.80\columnwidth]{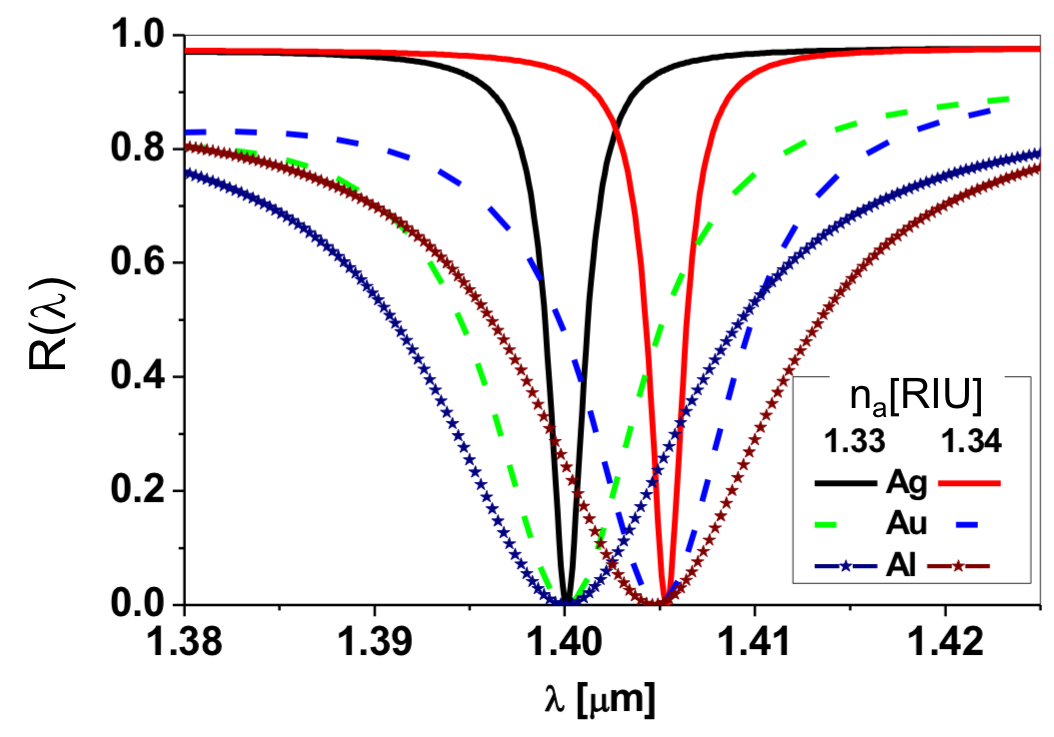}
  \caption{Spectral reflectance, $R(\lambda)$, of the optimized geometry for the three metals. They  are analyzed for two values of the index of refraction of the analyte: $n_a= 1.33$, and $n_a=1.34$. The spectral locations of the minima are redshifted as $n_a$ increases.}
  \label{fig:response2}
\end{figure}
  
For optimum performance, we find the following maximum values for the three metals under analysis: silver, 
MF$_{\rm Ag}$=108 at GW$_{\rm Ag}$=610 nm and GH$_{\rm Ag}$=1050 nm; 
gold, 
MF$_{\rm Au}$=53 at GW$_{\rm Au}$=750 nm and GH$_{\rm Au}$=830 nm; 
and aluminum, MF$_{\rm Al}$=40 when GW$_{\rm Al}$=850 nm and GH$_{\rm Al}$=770 nm. 
The white circles in Fig. \ref{fig:optimzation}  determine the location in the map where the merit function reaches the maximum value. 
Also, a combination of parameters close to those circles shows a large merit function that do not compromise the performance of the system. 

Figure \ref{fig:response2} shows the spectral reflectance of the structure made with the geometrical parameters maximizing the merit function. We observe here how the  response is narrower and slightly deeper for Ag (as expected from the values of the MF). The reflectances have a similar lineshape but they differ in their linewidth. The resonances  can be characterized by their quality factor parameter, $Q=\lambda_{\rm res}/\Delta \lambda$, where $\Delta \lambda$ can be given as the FWHM. Actually, the linewidth depends on the material parameters as it corresponds with their plasmonic resonance origin.  Using the definition for $Q$, we find the following values: $Q_{\rm Ag}=700$, $Q_{\rm Au}=155$ and $Q_{\rm Al}=88$. 
When changing the index of refraction, reflectivity remains almost invariant in shape, but is red-shifted when the index of refraction increases from 1.33 to 1.34.  A closer look to the spectral location of the minimum in reflectance (see Fig. \ref{fig:response2}) shows how  Ag reflectance moves a little farther to longer wavelengths than Au and Al responses. 
From our results, we conclude that a silver film  optimizes the device in terms of $Q$ and spectral shift. 
Furthermore, as long as the metal is not in direct contact with the analyte, biocompatibility issues of silver are overcome. 

\begin{figure}[h!]
\centering
  \includegraphics[width=0.950\columnwidth]{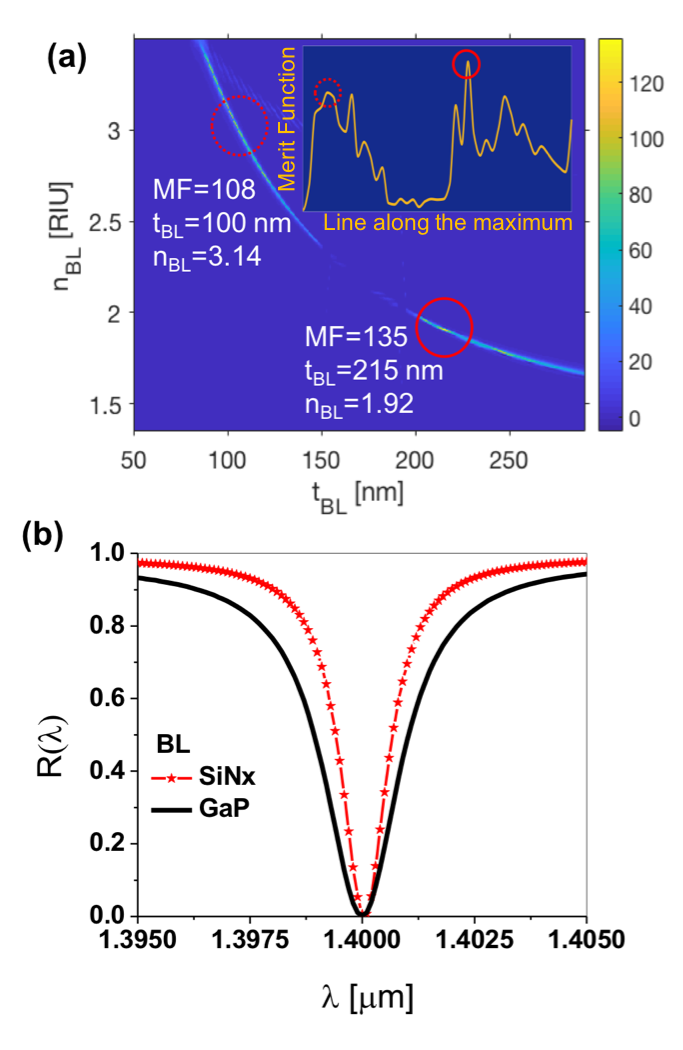}
  \caption{(a) Dependence of the merit function in terms of the characteristics of the buffer layer ($n_{\rm BL}, t_{\rm BL}$). Although the index of refraction cannot be varied continuously we may check that a higher index of refraction means a lower thickness of the buffer layer. The red circles locate the non-optimum (dashed) and optimum cases (solid). The inset shows the variation of the merit function (MF) along the line of the maximum values. (b) Spectral response of the device for a non-optimum buffer layer made of GaP (black) and the optimum case corresponding to buffer layer made of SiNx, having an index of refraction close to 1.92 (red).  
  }
  \label{fig:buffer}
\end{figure}

So far, we have optimized the proposed device by changing the geometric parameters of the grating and the material of the metal layer. Now, we consider the buffer layer and how its thickness (see $t_{\rm BL}$ in Fig. \ref{fig:gratings}), 
and index of refraction, $n_{\rm BL}$, allow a second round of optimization. The buffer layer is responsible for guiding the diffracted light towards the metal/dielectric interface (where SPRs are generated). Its index of refraction will affect to the effective index (effective dielectric permittivity, $\epsilon_{\rm eff}$) appearing in the wavevector matching condition (Eq. (\ref{eq:lambres})). The results of this optimization step are presented in Fig. \ref{fig:buffer} where we again find an optimum configuration within a range of feasible values. 
Although the starting point of the optimization was a GaP layer of thickness  100 nm (MF=108), we can see that the merit function is higher (MF=135) when the index of refraction is close to 2, and the thickness is around $t_{\rm BL}=215$ nm. The inset in Fig. \ref{fig:buffer}.a shows how the merit function varies along the line where the maximum values are located. A material well positioned to play as buffer layer is SiN$_x$  \cite{duttagupta_sinx_energy_2012}.
Fig. \ref{fig:buffer}.b shows the spectral reflectivity of the resonance for GaP and SiN$_x$. Also,
the material almost complying with the desired value in the index of refraction is Si$_3$N$_4$ (which is the same one used to 
build the grating nano-triangles). This is  a favorable situation  from the fabrication point of view because it saves an additional material process, and also  eliminates the index step between the nano-triangles and the buffer layer. The actual shape of the resonant minimum using Si$_4$N$_3$ is quite close to the one for SiN$_x$ represented in Fig. \ref{fig:buffer}.b.

\begin{figure}[h!]
\centering
  \includegraphics[width=0.96\columnwidth]{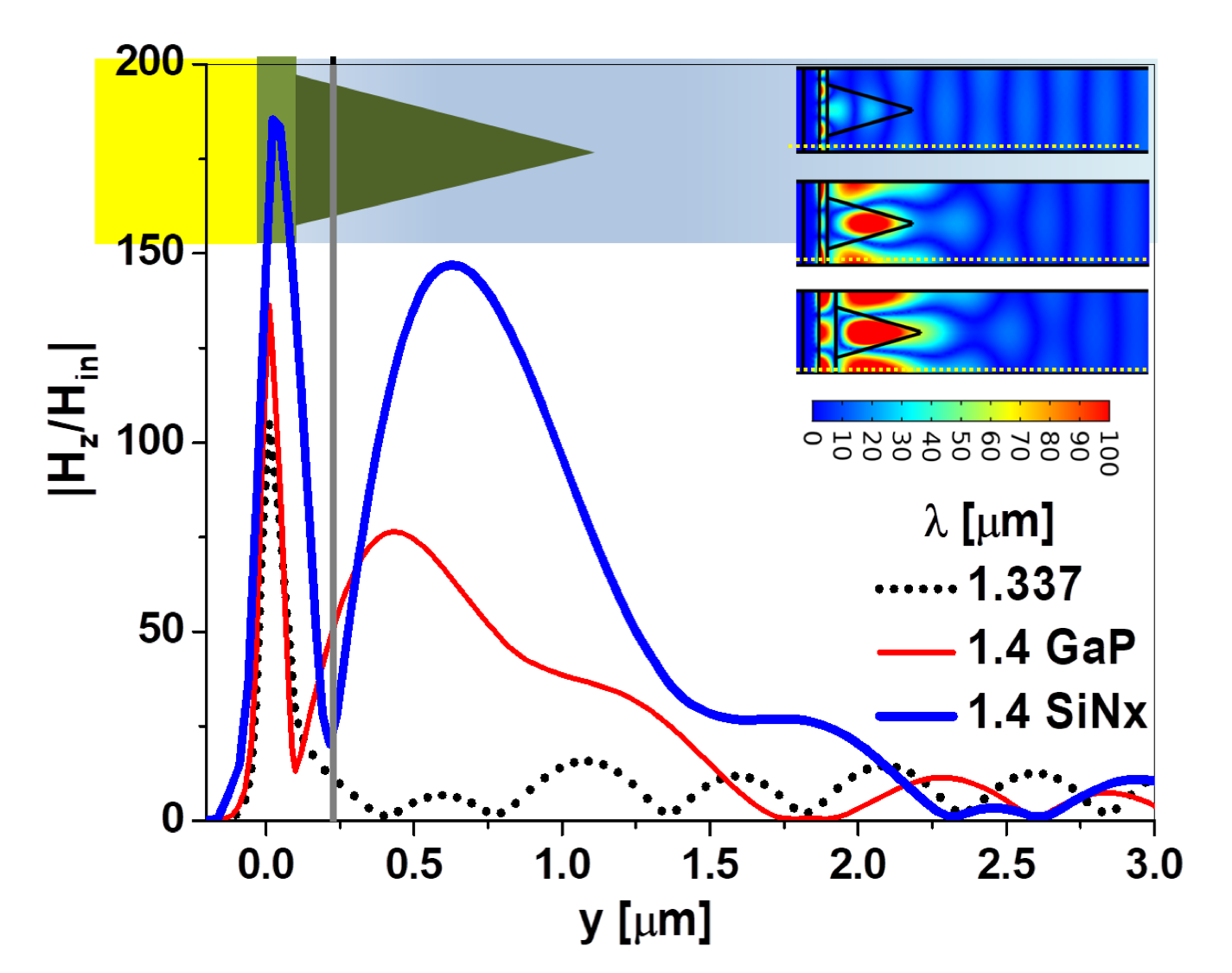}
  \caption{
Evolution of the $z-$component of the magnetic field, given as the ratio $|H(y)/H_{\rm in}|$,
as a function of the distance from the metal/dielectric interface ($y=0 \mu$m). Light is coming from the right. The dashed black line corresponds to $\lambda=1.337 \mu$m, and the solid red line is for $\lambda=1.4 \mu$m, both for a non optimized GaP grating material. The blue solid line is also for $\lambda=1.4 \mu$m, when the material of the grating and the buffer layer is the optimun one (SiN$_x$). The optimization also requires a thicker buffer layer of 200 nm, that is represented as a vertical gray solid line. This plot shows how resonances extend differently within the analyte medium. 
The insets represent the maps of the modulus of the magnetic field for the three cases. The dashed white lines in these maps locate the lines where the profiles are plotted.
}
  \label{fig:fields}
\end{figure}

The excited surface plasmon extends from the metal/dielectric surface towards the analyte and its amplitude decays inside the dielectric medium in contact with the metal. 
The buffer layer is made of GaP and it is 100 nm thick, which is thin enough since the surface plasmon amplitude  usually extends several hundreds of nanometers before completely decaying within the dielectric medium. 
Therefore, the electromagnetic field associated with the plasmon resonance interacts both with the grating material and the analyte.
Figure \ref{fig:fields} presents the evolution of the FE parameter for both resonances at $\lambda=1.337  \mu$m (in black) and $\lambda=1.4 \mu$m. In the case of $\lambda=1.4 \mu$m we have plotted the non-optimized case (in red) when the grating is made of GaP, and the optimized one (in blue) corresponding with a grating and buffer layer made of SiN$_x$.
 If we check how the field evolves from the metal/dielectric interface (see insets in Fig. \ref{fig:fields}) along the dashed lines in the map, for the case of 
 $\lambda=1.4\mu$m, we identify two maxima (see red and blue lines in Fig. \ref{fig:fields}). One of the maxima is located at the metal/dielectric surface, and the other is placed at the grating region.  Again, the resonance at $\lambda=1.4 \mu$m is better suited for its use in a sensor, because it interacts more intensely with the analyte medium at the grating gap. 
Actually, the field enhancement is as high as FE=108 for $\lambda=1.4 \mu$m, 4.7 times  larger than for $\lambda=1.337 \mu$m at the analyte volume.
Besides the field enhancement, the resonance extends deeper in the analyte at $\lambda=1.4 \mu$m, as predicted by the value of $h_{\rm eff}$ calculated through an effective index mixing model in subsection \ref{sec:proposedstructure}.

\section{Analysis and characterization}
\label{sec:deviceanalysis}

\begin{figure}[h!]
\centering
  \includegraphics[width=1.0\columnwidth]{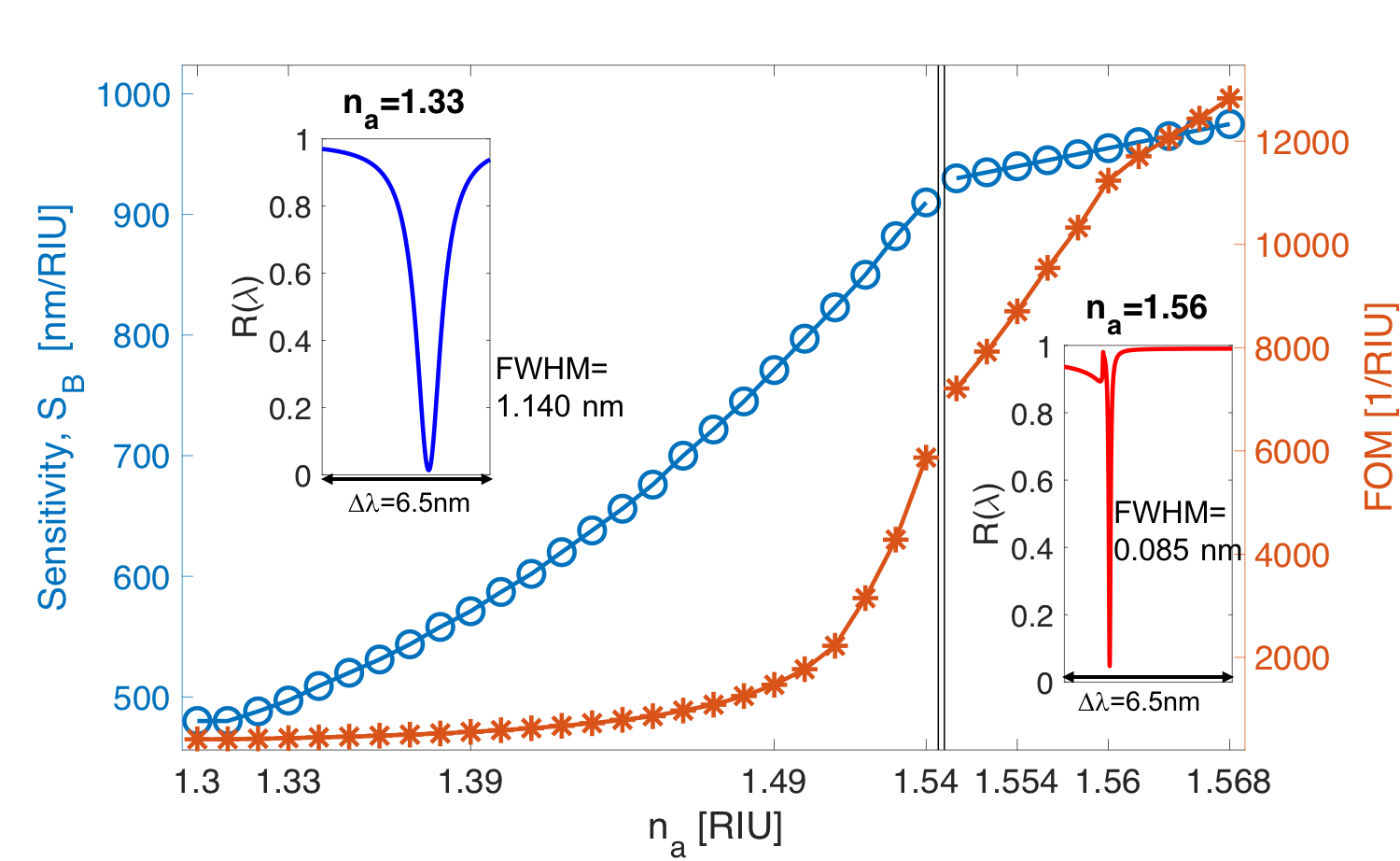}
  \caption{Sensitivity, and FOM for the optimized device at different values of analyte refractive index, $n_a$. The plot is divided into two portions. The left part describes the evolution from $n_a=1.3$ up to 
  $n_a=1.54$, and the right portion is for a narrower range: $n_a \in [1.55, 1.568]$. The insets show the lineshape of the resonance for $n_a=1.33$ (left), and $n_a=1.56$ (right) and the values of their FWHM. For better comparison, the plotted wavelength interval is the same (6.5 nm) for both insets. 
  }
  \label{fig:sens2}
\end{figure}

All the previous changes and parameterizations are used to evaluate sensitivity and FOM when these devices work as refractometric sensors \cite{Elshorbagy2017}. The definition of sensitivity is
\begin{equation}
S_{B}=\frac{\delta\lambda}{\delta n}
,\label{eq:sensi}
\end{equation}
and the figure of merit, FOM, is defined as
\begin{equation}
{\rm FOM}=\frac{S_{B}}{{\rm FWHM}}
,\label{eq:FOM}
\end{equation}
where $\delta\lambda$ is the wavelength shift of the resonance when the index of refraction of the analyte changes by an amount $\delta n$, and FWHM is the full width at half maximum of the resonance.
Therefore, to complete the evaluation of the device we  calculate the sensitivity and FOM for different analyte refractive index values, where the device operates efficiently, which determines the device dynamic range.
The geometric and material parameters of this optimum design are presented in table \ref{tab:optimumparameters}.
 Figure \ref{fig:sens2} represents how sensitivity and FOM change with the index of refraction of the analyte, $n_a$, within the studied index range. 
Sensitivity, $S_B$, is steadily increasing with $n_a$ from 480 nm/RIU (at $n_a=1.3$) till 975 nm/RIU at $n_a=1.568$. Also FOM increases with $n_a$, and shows  a large variation from 413 RIU$^{-1}$ (at $n_a=1.30$) until reaching a value of 12829 RIU$^{-1}$ (at $n_a=1.568$). This  change is caused by the abrupt narrowing of the spectral reflectance, which FWHM is as small as 0.08 nm for the resonance observed at $n_a=1.568$.  The calculation has stopped at $n_a=1.568$ because we consider the associated FWHM narrow enough to require a high-resolution spectrometer to operate the system.
These values are very competitive with current reported values of sensitivity and FOM for similar refractometric sensors, and show how the proposed nanostructure helps to improve the performance.
The insets in this figure represent the lineshape of the resonance for $n_a=1.33$ at the left, and $n_a=1.56$ to the right. These insets expand along the same wavelength interval of 6.5 nm, but the central wavelength of the resonance changes ($\lambda_{\rm res} (n_a=1.33) = 1.4001 \mu$m, and $\lambda_{\rm res} (n_a=1.56) =1.5603 \mu$m).

\begin{table}[h!]
\begin{center}
\caption{Geometrical parameters and materials for the optimum design
at $\lambda=1.4\mu{\rm m}$.
  \label{tab:optimumparameters}}
\begin{tabular}{ll}
\hline  
Material &  Dimensions  
\\  \hline 
SiO$_2$ (substrate)  & $\infty$ 
\\ \hline 
Ag (metal)  & $t_m=200$ nm  
\\ \hline  
SiN$_x$ (buffer layer)  & $t_{\rm BL}= 215$ nm 
\\ \hline 
Si$_3$N$_4$ (nanotriangle)  & Width, GW=610 nm    \\
			  & Height, GH = 1050 nm  \\
 			  & Period, $P$ = 1000 nm    
\\ \hline 
H$_2$O (analyte) & $\infty$    
\\ \hline 
\end{tabular}
\end{center}
\end{table}

At this point, we guess that the thin dielectric buffer layer, isolating the metal from the analyte, reduces the extent of the plasmonic resonance within the analyte volume. 
However, we have checked that by removing this isolation layer, the performance of the system is not compromised by this buffer layer because it increases the contrast of the index of refraction.
In fact, although a configuration without a buffer layer behaves slightly better for low index of refraction, the maximum sensitivity and FOM for the studied range of the index of refraction of the analyte, $n_a$, are worse ($S_{\rm B}= 867$ nm/RIU, and ${\rm FOM} = 5385 $ RIU$^{-1}$) than when including an optimized buffer layer.  
Moreover, this layer preserves the analyte from its contamination with the metallic layer, as Ag is highly keen to oxidation processes and hardly biocompatible.

\section{Conclusions}
\label{sec:conclusions}

In this paper we present a customizable nano-triangular structure on top of a flat dielectric/metal interface with tunable properties and  performance. It works as a refractometric sensor based on surface plasmon resonances. The low-order grating modes divert light towards the dielectric/metal interface fulfilling the wavevector matching conditions. 
We can tune the physical mechanism of this interaction through a careful evaluation of an effective index of refraction that depends on the material volumetric fraction where the wave accompanying the plasmon propagates. This effective index includes the analyte medium. 

The proposed configuration has been optimized for both the geometry and material selection. The materials that perform best are Ag for the metal, and Si$_3$N$_4$ for the  high-index material of the buffer layer and the nano-triangles.
This buffer layer plays two important roles: (i) it coats the silver surfaces enhancing device's biocompatibility, and (ii)  it provides a high contrast of the index of refraction that favors the existence of  field enhancement regions localized at the analyte volume. 

Once optimized, the device has been characterized as a refractometric sensor. Its dynamic range goes from 1.30 to 1.55 in the index of refraction of the analyte. 
This allows the device to operate at several wavelengths and makes multifunctional operation possible.
 The values of sensitivity ($S_B=940$ nm/RIU) are competitive with other proposed structures, plus the FOM sharply increases when moving to higher index of refraction because the spectral reflectivity minimum becomes narrower  as $n_a$ increases.

To summarize, we have demonstrated an efficient refractometric sensor that  
exploits the hybridization of grating modes and SPR. 
Its performance is highly competitive and it is also biocompatible.
We have considered current limitations in fabrication methods for the dimensions of the device.
Moreover, its capability to be excited under normal incidence conditions allows  its integration with optical fibers. 
We foresee a biomedical application: determination of tear samples  (involving nano or picoliters volumes) to improve the diagnose of dry-eye syndromes.

\section*{Acknowledgements}
This work has been partially supported by several sources:   
Ministerio de Econom\'{\i}a y Competitividad of Spain through projects TEC2016-77242-C3-1-R and DPI2016-75272-R; and Comunidad de Madrid SINFOTON2-CM (S2018/NMT-4326). These grants are also co-founded by the European Fund for Regional Development. ME is also supported by  Ministry of Higher Education of Egypt  (missions section).
The authors thanks I. Alda for her critical and careful reading of the manuscript, and the fruitful discussion along the development of this contribution.



\end{document}